\documentstyle[sprocl]{article}

\newcommand{\nn}{\nonumber}
\newcommand{\bd}{\begin{document}}
\newcommand{\ed}{\end{document}}
\newcommand{\bc}{\begin{center}}
\newcommand{\ec}{\end{center}}
\newcommand{\be}{\begin{eqnarray}}
\newcommand{\ee}{\end{eqnarray}}
\newcommand{\bt}{\begin{tabbing}}
\newcommand{\et}{\end{tabbing}}
\newcommand{\eqn}{\global\def\theequation}
\newcommand{\sw}{sin^2 \theta_W}
\newcommand{\fbd}{f_B}
\newcommand{\bi}{\bibitem}

\begin{document}

\title{ANOMALY, CHARGE QUANTIZATION AND FAMILY}

\author{C.~Q.~GENG}

\address{Department of Physics, National Tsing Hua University,
Hsinchu, Taiwan}


\maketitle\abstracts{ 
We first review the three known chiral anomalies
in four dimensions and then use the anomaly free
conditions to study the uniqueness of quark and lepton
representations and charge quantizations in the standard model.
We also extend our results to theory with an arbitrary number of
color. Finally, we discuss the family problem.}

 Although the standard model \cite{sd} of $SU(3)_C\times SU(2)_L\times
U(1)_Y$ has been remarkably successful experimentally, there are several
puzzles, such as why the electric charges of quarks and leptons are
quantized and why there are three fermion families? 
In this talk I would like to study these two puzzles in the viewpoint
of the chiral gauge anomaly cancellations. 

It is well-known that the anomaly free conditions
arising from the theoretical requirements of renormalizability and
self-consistency are the most elegant tool to test the gauge theory. Three
anomalies thus far have been identified for chiral gauge theories in
four dimensions: (1) The triangular (perturbative) chiral gauge anomaly,$^2$
which must be canceled to avoid the breakdown of gauge invariance and
renormalizability of the theory; we call this the {\it triangular}
anomaly. (2) The global (non-perturbative) $SU(2)$ chiral gauge anomaly,$^3$
which must be absent in order to define the fermion integral in a gauge
invariant way; we call this the $global$ anomaly. This anomaly was first
pointed out by Witten,$^3$ and is known as the Witten $SU(2)$ anomaly.
He showed in 1982 that any $SU(2)$ gauge
theory with an odd number of left-handed fermion (Weyl) doublets is
mathematically inconsistent. (3) The mixed (perturbative) chiral
gauge-gravitational anomaly,$^{4,5}$ which must be canceled in order to
ensure general covariance of the theory; we call this the $mixed$ anomaly.
This anomaly was first discussed by Delburgo and Salam$^4$ in 1972 and its
consequences studied by Alvarez-Gaum\'{e} and Witten$^5$ in 1983, who
concluded that a necessary
condition for consistency of the theory coupled to gravity is that the sum
of the $U(1)$ charges of the left-handed fermions vanishes, i.e., $TrQ=0$.

We now review
the three chiral anomalies for the simple Lie groups.

 1. {\em The Triangular Anomaly}. It has been shown$^6$ that the simple
Lie groups:
$SU(2)$, $SO(2k+1)(k>2)$, $SO(4k)(k\geq 2)$, $SO(4k+2)(k\geq 2)$,
$Sp(2k)$, $G_2$, $F_4$,
$E_6$, $E_7$, and $E_8$ are safe groups. The only simple groups with
possible
triangular anomaly are the unitary groups $SU(n)(n\geq 3)$. Therefore, if we
start with the groups which do not contain $SU(n)(n\geq 3)$ group, the theory
will be free of triangular anomaly.

2. {\em The Global Anomaly}. We classify the simple Lie groups $G$ into the
following two classes.
(I) $Sp(2k) (Sp(2)\simeq SU(2))$.
These groups$^{7}$ have the property of
\begin{eqnarray}
{\bf \Pi}_4(Sp(2k))={\bf Z}_2\ ,   
\end{eqnarray}
where ${\bf \Pi}_4$ is the fourth homotopy group and ${\bf Z}_2$ is the two-valued
discrete group (like parity). According to Witten,$^3$ the group $G^{(I)}=Sp(2k)$ 
has global anomaly if the number of fermion zero modes (for $SU(2)$ group,
it is equal to the number of fermion doublets) is odd.
(II) $SU(n)(n\geq 3), SO(2k+1)(k>2), SO(4k)(k\geq 2), SO(4k+2)(k\geq 2),
G_2, F_4, E_6, E_7,$ and $E_8$.
These groups ($G^{(II)}$) have no global anomaly since their fourth homotopy 
groups are trivial,$^{3,7}$ i.e.,
\begin{eqnarray}
{\bf \Pi}_4(G^{(II)})=0\ .   
\end{eqnarray}
 However, the interesting question$^{8}$ arises as to how one can know at
the level of $G^{(II)}$ whether such a theory is global anomaly-free when
$G^{(II)}$ breaks down to groups which contain $G^{(I)}$. Recently, we
present a sufficient condition$^{8}$ that for any simple group $G$,
containing $Sp(2k)$ as a subgroup, and for which ${\bf \Pi}_4(G)=0$, the
vanishing of the triangular
perturbative anomaly for Weyl representations of $G$ will guarantee the 
absence of the global non-perturbative $Sp(2k)$ anomaly.

3. {\em The Mixed Anomaly}.
This anomaly is non-trivial only for the theory in which there is $U(1)$ 
symmetry with non-zero total charges.$^5$ Obviously, all the simple Lie groups
($G^{(I),(II)}$) are safe groups. Furthermore, when these groups break
down to groups which contain $U(1)$, e.g.,
\begin{eqnarray}
G \rightarrow g\times \prod_i U(1)_i\ ,
\end{eqnarray}
unlike the previous case, there is no mixed anomaly since the $U(1)$ 
operators are the generators of $G$ and must be traceless. 

The triangular anomaly-free of the standard model was first noted$^9$ in
1972 for each quark-lepton family. It was clear that with only the
triangular anomaly-free condition$^{10}$ one could not explain the
empirically determined quark-lepton representations and their quantized
hypercharges.
We now study$^{11}$ the question of the uniqueness of
quarks and leptons in the standard model by insisting on all three anomaly-free
conditions. With an arbitrary color number $N\ (\ge 3)$, we begin by
allowing an arbitrary number of (left-handed) Weyl
representations under the group of
$SU(N)\times  SU(2) \times U(1)$, i.e.,

\be
SU(N)\times  SU(2) \times U(1)&&
\nn\\
N~~~~~~~~~2~~~~~~~~~Q_i\,,&&  i=1,\cdots, j
\nn\\
N~~~~~~~~~1~~~~~~~~~Q_i'\,,&& i=1,\cdots, k\nn\\
\overline{N}~~~~~~~~~1~~~~~~~~~\overline{Q}_i\,,&& i=1,\cdots, l
\\
\overline{N}~~~~~~~~~2~~~~~~~~~\overline{Q}_i'\,,&& i=1,\cdots, m
\nn\\
1~~~~~~~~~2~~~~~~~~~q_i\,,\ &&i=1,\cdots, n
\nn\\
1~~~~~~~~~1~~~~~~~~~\overline{q}_i\,,\ &&i=1,\cdots, p
\nn
\ee
where the integers $j,k,l,m,n$ and $p$ and the $U(1)$ charges
are all arbitrary. The triangular anomaly free conditions then lead to
the following equations:
{\small
\be
[SU(N)]^3&:&\sum_{i=1}^j2 +\sum_{i=1}^k2-
\sum_{i=1}^l1-\sum_{i=1}^m2 =0\ ,
\nn\\
{[SU(N)]^2}\,U(1)&:& 2 \sum_{i=1}^jQ_i +\sum_{i=1}^kQ_i'+
\sum_{i=1}^l\overline{Q}_i' +2\sum_{i=1}^m\overline{Q}_i'=0\ ,\\
{[SU(2)]^2}\,U(1)&:&N\sum_{i=1}^jQ_i +N\sum_{i=1}^m\overline{Q}_i'
+\sum_{i=1}^nq_i=0\ , \nn\\
{[U(1)]^3}:N\sum_{i=1}^jQ_i^3&+&{N\over 2}\sum_{i=1}^kQ_i'^3+
{N\over 2}\sum_{i=1}^l\overline{Q}_i^3+N\sum_{i=1}^m\overline{Q}_i'^3+
\sum_{i=1}^nq_i^3+{1\over 2}\sum_{i=1}^p\overline{q}_i^3=0\ .\nonumber
\ee }
The global $SU(2)$ anomaly-free condition is
\begin{eqnarray}           
N\,j+N\,m+n=E\ ,
\end{eqnarray}
\\
where $E$ is an even integer. Finally the mixed anomaly-free
condition is 
{\small
\begin{eqnarray}           
[U(1)]: N\sum_{i=1}^jQ_i+{N\over 2}\sum_{i=1}^kQ_i'+
{N\over 2}\sum_{i=1}^l\overline{Q}_i+N\sum_{i=1}^m\overline{Q}_i'+
\sum_{i=1}^nq_i+{1\over 2}\sum_{i=1}^p\overline{q}_i=0\ .
\end{eqnarray} }

The requirements of minimality and the three anomaly-free conditions
[Eqs.\ (5)-(7)] lead to the values: 
(I) if N= even \#, $j=1$, $k=0$, $l=2$, $m=n=p=0$, and 
\begin{eqnarray}           
Q_1=0\,,\
\overline{Q}_1=- \overline{Q}_2\,;
\end{eqnarray}           
and 
(II) if N= odd \#, $j=1$, $k=0$, $l=2$, $m=0$, $n=1$, $p=1$, and
to two solutions of $U(1)$ charges 
\begin{eqnarray}           
&&Q_1={1\over N}\,,\ \overline{Q}_1=
-{N+1\over N}\,,\
\overline{Q}_2=
{N-1\over N}\,,\ 
\overline{q}_1=-2q_1=-2,,
\\
&&Q_1=q_1= \overline{q}_1=0\,,\
\overline{Q}_1=- \overline{Q}_2\,,
\end{eqnarray}           
where we have chosen the normalization $q_1=-1$ in Eq. (9).
For $N=3$, the solutions in Eqs. (9) and (10) are the
``standard model'' and the so called ``bizarre'' ones, respectively.
We note that the ``inert'' state (1,1,0) for the ``bizarre'' solution
is a non-chiral representation
and it must be excluded.
It is interesting to note that the ``bizarre'' solution may be viewed as 
the standard one when $N\rightarrow \infty$. Without
considering the ``bizarre'' solution, for the odd number of color,
all the $U(1)$ charges are uniquely determined.
In this case,
the resulting Weyl representations of $SU(N)$ and $SU(2)$ and
their $U(1)$ charges 
are those in the standard model if N=3 (cf.\ Table 1).
The electric charges of quarks and leptons 
for an arbitrary odd number of color $N$ are given in Table 1
where the electroweak symmetry is spontaneously broken down to
$U(1)_{EM}$ by the Higgs mechanism.

\begin{center}
{\bf Table 1.}\  The quantum numbers of quark and lepton representations
under $SU(N)_C\times SU(2)_L\times U(1)_Y$ and $SU(N)_C\times U(1)_{EM}$
\\
\vspace{0.5cm}

\renewcommand{\arraystretch}{2}
\noindent\begin{tabular*}{11cm}{@{\extracolsep{\fill}}cccc}
\hline \hline
Particles  & $SU(N)_C\times SU(2)_L\times U(1)_Y$& $\rightarrow$&
$SU(N)_C\times U(1)_{EM}$\\
$(i=1,2,3)$ &&&\\
\hline
$\displaystyle{u\choose d}_{\!L}^i$ & N~~~~~~~~~~~2~~~~~~~~~
$\phantom{-}\displaystyle{1\over N}$ && $ 
\displaystyle\left({N\atop N}\right.$~~~~
$\displaystyle\left.{\phantom{-}{N+1\over 2N}\atop -{N-1\over 
2N}}\right)$\\
${u^c_L}^i$ & $\overline{N}$~~~~~~~~~~~1~~~~~$-\displaystyle{N+1\over
N}$ &&$\overline N$~~~~~~~$-{N+1\over 2N}$\\
${d^c_L}^i$ & $\overline{N}$~~~~~~~~~~~1~~~~~$
\phantom{-}\displaystyle{N-1\over N}$&&
$\overline N$~~~~~~~$\phantom{-}{N-1\over
2N}$  \\
$\displaystyle{\nu\choose e}_{\!L}^i$ &1~~~~~~~~~~~~2~~~~~~~~~~$-1$&&
$ \displaystyle\left({1\atop 1}\right.$~~~~~~~
$\displaystyle\left.{\phantom{-}0\atop -1}\right)$\\
${e^c_L}^i$
&1~~~~~~~~~~~~1~~~~~~~~~~$\phantom{-}2$&&\phantom{-}1~~~~~~~~~~~~\phantom{-}1\\
\hline \hline
\end{tabular*}
\end{center}

For the standard model of $SU(3)_C\times SU(2)_L\times U(1)_Y$,
we thus find that the requirements of minimality
and freedom from all three 
chiral gauge anomalies lead to a unique set of Weyl representations (and
their $U(1)_Y$ charges) of the standard group that correspond to the observed
quarks and leptons of one family. Furthermore, the $U(1)_Y$ charges of these
quarks and leptons are quantized and correctly determined by adding the mixed
anomaly-free condition and thus a long-standing puzzle of the electric charge
quantization of quark and lepton can be solved within the content of the
standard model.

In spite of the success of the standard model, it is still a mystery why the
three anomaly cancellations, especially the global and the mixed ones,
should be satisfied. Naturally one hopes that new physics beyond the
standard model can 
provide us an explanation to this question. 
 From the above studies we see that the three anomaly-free conditions in
the standard model may be automatically satisfied if it comes from a large
group, especially,
a grand unification group. For example, with the $E_6$ grand unification 
theory, the triangular, the global, and the mixed anomalies are trivial at the
level of $E_6$ which guarantees their freedom at the standard group level. We 
thus conclude that the resolution of the question of the uniqueness of
the massless
fermion representations and $U(1)_Y$ charges for the standard group --
when viewed
from the standpoint of the perturbative triangular and mixed chiral
gauge-gravitational anomalies and the absence of the 
non-perturbative global $SU(2)$ chiral gauge anomaly in four dimensions --
argues strongly for some new physics beyond the standard model.

Finally, we discuss the family issue.
It is clear that, as one can see from the above study, the
imposition of all three anomaly-free
conditions for the standard model does not shed any immediate light on the
``generation problem''. In fact, the quantum numbers in Table 1 are
generation blind.
Moreover, if one enlarges the standard group to include an $SU(2)$
or $SU(3)$ group, one can show that the theories are precisely the one family
fermion structure of the left-right symmetric model$^{11}$ $SU(3)_C\times
SU(2)_L\times SU(2)_R\times U(1)$ and the chiral-color model,$^{12}$
$SU(3)_{CL}\times SU(3)_{CR}\times SU(2)_L\times U(1)_Y$, respectively, instead
of having a family group.  
Clearly, some new ideas$^{13}$ 
are needed to constrain on
the number of families which would be a key to the new physics.
We now present a toy model which gives rise to three families of quarks
and leptons. 
In the standard model, in each family there are 15 Weyl spinors.
With a right handed neutrino, the number becomes 16.
For three families, the total numbers are 48.  One may put all these 48
Weyl spinors into a flavor box to form a large global symmetry as
$U(48)$.$^{13}$
 From the study in Eqs. (4)-(10), we can extend the group
of $SU(N)\times SU(2)\times U(1)$ with both even and odd numbers
of $N$ to a larger group of $SU(N)\times SU(2)\times SU(2)$
in which $N$ has to be an even number
as shown in Table 2. For N=4, it is just
the Pati-Salam model,$^{14}$ which contains a right-handed neutrino.
We remark that the representations under $SU(N)\times SU(2)\times SU(2)$
in Table 2 are unique unlike the case with a $U(1)$ symmetry and there is
no more ``bizarre'' solution like the one in Eq. (10).

\begin{center}
{\bf Table 2.}\  The fermion quantum numbers
under $SU(N)\times SU(2)\times SU(2)$
\\
\vspace{0.5cm}

\renewcommand{\arraystretch}{2}
\noindent\begin{tabular*}{6cm}{@{\extracolsep{\fill}}ccccc}
\hline \hline
$SU(N)_C$&$\times$& $SU(2)$&$\times$ &$SU(2)$\\
\hline
$N$&&2&&1\\
$\overline{N}$&&1&&2\\
\hline \hline
\end{tabular*}
\end{center}
\vspace{.5cm}

 We now take the global flavor symmetry $U(48)$ and gauge its subgroup
$SU(12)\times SU(2)\times SU(2)$ so that the fermions transform according
to the representations given in Table 2 with $N=12$.
Thus, the model is a generalized Pati-Salam theory with the color
being 12. The symmetry breaking chains by various suitable scalars are
given as follows:

\bc
$SU(12)_C\times SU(2)_L\times SU(2)_R$\\
           $12\ \ \ \ \ \ \ \ \ \ \ \ 2\ \ \ \ \ \ \ \ \ \ \ 1$\\
$\overline{12}\ \ \ \ \ \ \ \ \ \ \ \ 1\ \ \ \ \ \ \ \ \ \ \ 2$ \\
$\downarrow$ \\
$SU(12)_C\times SU(2)_L\times SU(2)_R$\\
$\downarrow$ \\
$SU(8)_C\times SU(4)_{C1}\times SU(2)_L\times SU(2)_R\times U(1)$\\
$\downarrow$ \\
$SU(4)_{C3}\times SU(4)_{C2}\times SU(4)_{C1}\times SU(2)_L\times SU(2)_R
\times U(1)\times U(1)$\\
$\downarrow$ \\
$\downarrow$ \\
$SU(4)_{C}\times SU(2)_L\times SU(2)_R$\\
$\downarrow$ \\
$\downarrow$ \\
$SU(3)_{C}\times SU(2)_L\times U(1)_Y$\\
\vspace{.3cm}
$\underbrace{three\ quark\ and\ lepton\ families}$
\ec

\noindent
Therefore, there are three generations of quarks and leptons
under the standard group of $SU(3)_{C}\times SU(2)_L\times U(1)_Y$.
However, before taking this model seriously, more works have to be done.

In sum,
we have found that the requirements of minimality
and freedom from all three 
chiral gauge anomalies in four dimensions lead to a unique set of Weyl
representations 
of the standard group, 
corresponding to the observed quarks and leptons of one family.
Furthermore,
the $U(1)_Y$ charges of these quarks and leptons are quantized and
correctly determined by adding the mixed
anomaly-free condition and thus a long-standing puzzle of the electric charge
quantization of quark and lepton can be solved within the content of the
standard model.
The determination of the uniqueness of the standard model due to the
{anomaly cancellations} argues strongly for new physics beyond the
standard model, especially some form of the quark-lepton unification.
However, there is still no answer to the family problem. Maybe there are
possibly some as-yet-unidentified
anomalies in four dimensions, or  larger symmetries like
$SU(12)_C\times SU(2)_L\times SU(2)_R$, or higher dimensions,$^{13}$ or
presons,$^{15}$ or others.

\begin{center}
{\bf Acknowledgements}
\end{center}

This work was supported in part by the National Science Council of the
Republic of China under contract number NSC-89-2112-M-007-013.
  

\end{document}